\documentstyle[12pt]{article}
\title{An Algorithm for the Longest Common Subsequence and Substring Problem}
\author{ Rao Li \\
 Dept. of Computer Science, Engineering, and Mathematics \\
         University of South Carolina Aiken \\
	   Aiken, SC 29801 \\
          USA \\
         {\it Email: raol@usca.edu } \and
Jyotishmoy Deka \\
Dept. of Electrical Engineering \\
Tezpur University \\
Tezpur, Assam 784028 \\
India \\
{\it Email: jyotishmoydeka62@gmail.com}  \and 
 Kaushik Deka \\
Dept. of Computer Science and Engineering \\
National Institute of Technology Silchar \\
Cachar, Assam 788010 \\
India \\ 
{\it Email: jagatdeka20@gmail.com} 
}
\date{July 18, 2023}

\begin{document}
\maketitle
\newpage
\begin{abstract}
In this note, we first introduce a new problem called the longest common subsequence and substring problem. Let $X$ and $Y$ be two strings over an alphabet $\Sigma$.
The longest common subsequence and substring problem for $X$ and $Y$ is to find the longest string which is a subsequence of $X$ and a substring of $Y$. 
We propose an algorithm to solve the problem.    
 \end{abstract} 
$$Keywords: Algorithm, the \,\, longest \,\, common \,\,  subsequence, $$  
 \hspace*{37mm}  $the \,\, longest \,\, common \,\, substring$
\\

\noindent {\bf 1.  Introduction} \\

Let $\Sigma$ be an alphabet and $S$ a string over $\Sigma$. A subsequence of a string $S$ is obtained by deleting zero or more letters of $S$.
A substring of a string $S$ is a subsequence of $S$ consists of consecutive letters in $S$.
Let $X$ and $Y$ be two strings over an alphabet $\Sigma$.  The longest common subsequence problem for $X$ and $Y$ is to find the longest string which is a subsequence of both $X$ and $Y$. 
The longest common substring problem for $X$ and $Y$ is to find the longest string which is a substring of both $X$ and $Y$. Both the longest common subsequence problem and the longest common substring problem
have been well-studied in last several decades. They have applications in different fields, for example, in molecular biology, the lengths of the longest common subsequence and the longest common substring  
are the  suitable measurements for the similarity between two biological sequences. 
More details on the algorithms for the first problem
can be found in \cite{Apo}, \cite {Apo1}, \cite{Bergroth}, \cite{Cormen}, \cite{Hberg}, 
and \cite{Rick} and the second problem can be found in \cite{G} and \cite{W}.  Motivated by the two problems above, we introduce a new problem called the longest common subsequence and substring problem. 
The longest common subsequence and substring problem for $X$ and $Y$ is to find the longest string which is a subsequence of $X$ and a substring of $Y$. 
In this note, we propose an algorithm to solve this problem.    \newline

\noindent {\bf 2.  The Foundations of the Algorithm} \\

In order to present our algorithm, we need to prove some facts which are the foundations for our algorithm. Before proving the facts, we need some notations as follows.
For a given string $S = s_1 s_2 ... s_l$ over an alphabet $\Sigma$, the size of $S$, denoted $|S|$, is defined as the number of letters in $S$.
The $i$ prefix of $S$ is defined as $S_i = s_1 s_2 ... s_i$, where $1 \leq i \leq l$. Conventionally, $S_0$ is defined as an empty string.
The $l$ suffixes of $S$ are the strings of $s_1 s_2 ... s_l$, $s_2 s_3 ... s_l$, ..., $s_{l - 1}s_l$, and $s_l$. Let $X = x_1 x_2 ... x_m$ and $Y = y_1 y_2 ... y_n$ 
be two strings.  We define $Z[i, j]$ as a string satisfying the following conditions, where $1 \leq i \leq m$ and $1 \leq j \leq n$. \\  \\
 \hspace*{27mm} ($1$) It is a subsequence of $X_i$. \newline
 \hspace*{27mm} ($2$) It is a suffix of $Y_j$. \newline
 \hspace*{27mm} ($3$) Under ($1$) and ($2$), its length is as large as possible. \\

\noindent{\bf Fact 1.} Let $U = u_1 u_2 ... u_r$ be a longest string which is a subsequence of $X$ and substring of $Y$. Then $r = \max \{\, |Z[i, j]| : 1 \leq i \leq m,  1 \leq j \leq n \,\}$. \\

\noindent {\bf Proof of Fact 1.} For each $i$ with $1 \leq i \leq m$ and each $j$ with $1 \leq j \leq n$, we, from the definition of $Z[i, j]$, have that $Z[i, j]$ is a subsequence of $X$ and substring of $Y$.
By the definition of $U$, we have that $|Z[i, j]| \leq |U| = r$. Thus $\max \{\, |Z[i, j]| : 1 \leq i \leq m,  1 \leq j \leq n \,\} \leq r$. \\

Since $U = u_1 u_2 ... u_r$ is a longest string which is a subsequence of $X$ and a substring of $Y$, there is an index $s$ and an index $t$ such that $u_r = x_s$ and $u_r = y_t$ such that
$U = u_1 u_2 ... u_r$ is a subsequence of $X_s$ and a suffix of $Y_t$. From the definition of $Z[i, j]$, we have that $r \leq |Z[s, t]| \leq \max \{\, |Z[i, j]| : 1 \leq i \leq m,  1 \leq j \leq n \,\}$.  \\

Hence $r = \max \{\, |Z[i, j]| : 1 \leq i \leq m,  1 \leq j \leq n \,\}$ and the proof of Fact $1$ is complete. \\

\noindent{\bf Fact 2.} Suppose that $X_i = x_1 x_2 ... x_i$ and $Y_j = y_1 y_2 ... y_j$, where $1 \leq i \leq m$ and $1 \leq j \leq n$. 
If $Z[i, j] = z_1 z_2 ... z_a$ is a string satisfying conditions ($1$), ($2$), and ($3$) above. Then we have \newline

\noindent [1]. If $x_i = y_j $, then $a$ = 1 + the length of a longest string which is a subsequence  of $X_{i - 1}$ and a suffix of $Y_{j - 1}$. \newline
          
\noindent [2]. If $x_i \neq y_j$, then $a$ = the length of the longest string which is a subsequence of $X_{i - 1}$ and a suffix of $Y_{j }$. . \newline

\noindent {\bf Proof of [1] in Fact 2.} Suppose $W = w_1 w_2 ... w_b$ is a string satisfying the following conditions.  \\ \\
 \hspace*{27mm} ($1$) It is a subsequence of $X_{i - 1}$. \newline
 \hspace*{27mm} ($2$) It is a suffix of $Y_{j - 1}$. \newline
 \hspace*{27mm} ($3$) Under ($1$) and ($2$), its length is as large as possible. \\ 

\noindent  Since $W = w_1 w_2 ... w_b$ is  a subsequence of $X_{i - 1}$, a suffix of $Y_{j - 1}$, and $x_i = y_j $, 
$W = w_1 w_2 ... w_bx_i$ is  a subsequence of $X_i$ and a suffix of $Y_j$. From the definition of $Z[i, j]$, we have $|W| + 1 = b + 1 \leq |Z[i, j]| = a$.  \\

Since  $Z[i, j] = z_1 z_2 ... z_a$ is a string satisfying conditions ($1$), ($2$), and ($3$) above, we have that $z_a = y_j = x_i$. We further have that $z_1 z_2 ... z_{a - 1}$  is a string which is a subsequence of $X_{i - 1}$ and a suffix of  $Y_{j - 1}$. From the definition of
$W = w_1 w_2 ... w_b$, we have that $a - 1 \leq b$. Thus $a = 1 + b$ and $a$ = 1 + the length of a longest string which is a subsequence  of $X_{i - 1}$ and a suffix of $Y_{j - 1}$.  \\

\noindent {\bf Proof of [2] in Fact 2.} Suppose $U = u_1 u_2 ... u_c$ is a string satisfying the following conditions.  \\ \\
 \hspace*{27mm} ($1$) It is a subsequence of $X_{i - 1}$. \newline
 \hspace*{27mm} ($2$) It is a suffix of $Y_{j }$. \newline
 \hspace*{27mm} ($3$) Under ($1$) and ($2$), its length is as large as possible. \\ 

\noindent Since $U = u_1 u_2 ... u_c$ is  a subsequence of $X_{i - 1}$ and a suffix of $Y_{j }$,
$U = u_1 u_2 ... u_c$ is  a subsequence of $X_{i } $ and a suffix of $Y_j$. From the definition of $Z[i, j]$, we have $|U| = c \leq |Z[i, j]| = a$.  \\

Since $Z[i, j] = z_1 z_2 ... z_a$ is a string satisfying conditions ($1$), ($2$), and ($3$) above, we have that $z_a = y_j \neq x_i$. Thus  $z_1 z_2 ... z_{a}$  is a string which is a subsequence of $X_{i - 1}$ and a suffix of  $Y_{j}$. From the definition of
$U = u_1 u_2 ... u_c$, we have that $a  \leq c$. Thus $a = c$ and $a$ = the length of a longest string which is a subsequence  of $X_{i - 1}$ and a suffix of $Y_{j}$.  \\

Hence the proof of Fact $2$ is complete. \\

\noindent {\bf 3. An Algorithm for the Longest Common Subsequence and Substring Problem} \newline

Based on the Fact $1$ and Fact $2$ in Section$2$, we can design an algorithm for the longest common subsequence and substring problem. 
Once again, we assume that $X = x_1 x_2 ... x_m$ and $Y = y_1 y_2 ... y_n$ 
are two strings over an alphabet  $\Sigma$.
In the following Algorithm $A$, 
$W$ is an $(m + 1) \times (n + 1)$ array and
the cells $W(i, j)$, where $1 \leq i \leq m$ and $1 \leq j \leq n$, store the lengths of strings such that each of them satisfies the following conditions. \\  \\
 \hspace*{27mm} ($1$) It is a subsequence of $X_i$. \newline
 \hspace*{27mm} ($2$) It is a suffix of $Y_j$. \newline
 \hspace*{27mm} ($3$) Under ($1$) and ($2$), its length is as large as possible. \\ \\  

\noindent $ALG \, A (X, Y, m, n, W)$ \newline
$1.$ Initialization: $W(i, 0) \leftarrow 0$, where $i = 0, 1, ..., m$ \newline
\hspace*{29mm}      $W(0, j) \leftarrow 0$, where $j = 0, 1, ..., n$ \newline
 \hspace*{29mm}     $maxLength = 0$  \\
\hspace*{29mm}      $lastIndexOnY = n$ \\
\noindent $2.$ {\bf for} $i \leftarrow 1$ {\bf to} $m$  \newline
\noindent $3.$  \hspace*{8mm} {\bf for} $j \leftarrow 1$ {\bf to} $n$  \newline
\hspace*{22mm} {\bf if } $x_i = y_j$ \,  $W(i, j) \leftarrow W(i-1, j-1) + 1$ \newline
\hspace*{22mm} {\bf else} \, $W(i, j) \leftarrow W(i-1, j)$ \newline
\hspace*{22mm} {\bf if } $W(i, j) > maxLength$ \\
\hspace*{32mm}  $maxLength = W(i, j)$ \\
\hspace*{32mm}  $lastIndexOnY = j$ \\
\noindent $4.$ {\bf return} $A \,\, substring \,\, of \, Y \,\, between \,\, (lastIndexOnY \, - \, maxLength)$ \newline
\hspace*{19mm} $and \,\, lastIndexOnY$ \\

Because of the Fact $1$ and Fact $2$ in Section 2, it is clear that Algorithm $A$ is correct. Obviously, the time complexity of 
Algorithm $A$ is $O(mn)$ and the space complexity of Algorithm $A$ is also $O(mn)$.  \\

We implemented Algorithm A in Java and the program can be found at ``https://sciences.usca.edu/math/\textasciitilde mathdept/rli/LCSSeqSStr/LCSS.pdf".




\end{document}